\documentstyle[prl,twocolumn,aps,epsfig]{revtex}

\newcommand{\Lie}[0]{{\cal L}\, }

 \makeatletter
 \newcommand{\pback}[1]{{
   \let\@rrow=\leftarrowfill
   \mathchoice{\AIN@stemPullBack{#1}{\@rrow}}{\AIN@stemPullBack{#1}{\@rrow}}
     {\AIN@indxPullBack{#1}{\@rrow}}{\AIN@indxPullBack{#1}{\@rrow}}}
   \vphantom{#1}}

 \newcommand{\AIN@stemPullBack}[2]{
   \vtop{\mathsurround=0pt
   \ialign{##\crcr$\textstyle{#1}\strut$\crcr
     \noalign{\kern-0.4ex\nointerlineskip}{\tiny#2}\crcr}}}

 \newcommand{\AIN@indxPullBack}[2]{
   \vtop{\mathsurround=0pt
   \ialign{##\crcr\hfil$\scriptstyle{#1}$\hfil\crcr
     \noalign{\kern+0.4ex\nointerlineskip}{\tiny#2}\crcr}}}

\makeatother

\newcommand{\hateq}[0]{\mathrel{\widehat\mathalpha{=}}}
\newcommand{\emA}{{\mathbf{A}}}
\newcommand{\emF}{{\mathbf{F}}}

\def\l{\ell}
\def\ls{{(\l)}}
\def\G{{\cal G}_\Delta}

\def\bar{\overline}
\def\ba{\begin{eqnarray}}
\def\ea{\end{eqnarray}}
\def\be{\begin{equation}}
\def\ee{\end{equation}}
\def\={\hateq}
\def\puto#1{\rlap{\raise.5ex\hbox{\char'27}}{#1}}

\begin{document}
\draft \wideabs{\title{Generic Isolated Horizons and their
Applications}
\author{Abhay\ Ashtekar${}^1$, Christopher\ Beetle${}^{1}$, Olaf \
Dreyer${}^{1}$,\\ Stephen\ Fairhurst${}^1$, Badri\ Krishnan${}^{1}$,
Jerzy Lewandowski${}^{2,1}$ and Jacek Wi\'sniewski${}^{1}$}
\address{1. Physics Department, 104 Davey, Penn State, University
Park, PA 16802, USA\\ 2. Institute of Theoretical Physics, University
of Warsaw, ul. Ho\.{z}a 69, 00-681 Warsaw, Poland}

\maketitle

\begin{abstract}

Boundary conditions defining a generic isolated horizon are introduced.
They generalize the notion available in the existing literature by
allowing the horizon to have distortion and angular momentum.
Space-times containing a black hole, itself in equilibrium but possibly
surrounded by radiation, satisfy these conditions.  In spite of this
generality, the conditions have rich consequences.  They lead to a
framework, somewhat analogous to null infinity, for extracting physical
information, but now in the \textit{strong} field regions.  The
framework also generalizes the zeroth and first laws of black hole
mechanics to more realistic situations and sheds new light on the
`origin' of the first law.  Finally, it provides a point of departure
for black hole entropy calculations in non-perturbative quantum
gravity.

\end{abstract}
\pacs{Pacs: 04070-m, 0425Dm, 0460-m} }

A great deal of analytical work on black holes in general relativity
centers around event horizons in globally stationary space-times (see,
e.g., \cite{mh,w}).  While it is a natural starting point, this
idealization seems overly restrictive from a physical point of view. In
a realistic gravitational collapse, or a black hole merger, the final
black hole is expected to rapidly reach equilibrium. However, the
exterior space-time region will not be stationary.  Indeed, a primary
goal of many numerical simulations is to study radiation emitted in the
process.  Similarly, since event horizons can only be determined
retroactively after knowing the entire space-time evolution, they are
not directly useful in many situations. For example, when one speaks of
black holes in centers of galaxies, one does not refer to event
horizons. The idealization seems unsuitable also for black hole
mechanics and statistical mechanical calculations of entropy. Firstly,
in ordinary equilibrium statistical mechanics, one only assumes that
the system under consideration is stationary, not the whole universe.
Secondly, from quantum field theory in curved space-times,
thermodynamic considerations are known to apply also to cosmological
horizons\cite{gh}. Thus, it seems desirable to replace event horizons
by a quasi-local notion and develop a detailed framework tailored to
diverse applications, from numerical relativity to quantum gravity,
without the assumption of global stationarity. The purpose of this
letter is to present such a framework.

Specifically, we will provide a set of quasi-local boundary conditions
which define an isolated horizon $\Delta$ representing, for example,
the last stages of a collapse or a merger, and focus on space-time
regions admitting such horizons as an inner boundary. Although the
boundary conditions are motivated purely by geometric considerations,
they lead to a well-defined action principle and Hamiltonian framework.
This, in turn, leads to a definition of the horizon mass $M_\Delta$ and
angular momentum $J_\Delta$. These quantities refer only to structures
intrinsically available on $\Delta$, without any reference to infinity,
and yet lead to a generalization of the familiar laws of black hole
mechanics. We will also introduce invariantly defined coordinates near
$\Delta$ and a Bondi-type expansion of the metric. Finally, our present
boundary conditions allow distorted and rotating horizons and are thus
significantly weaker than those introduced in earlier papers
\cite{abf}. With this extension, the framework becomes a robust new
tool in the study of classical and quantum black holes.

For brevity, in the main discussion we will restrict ourselves to the
Einstein-Maxwell theory in four space-time dimensions.  Throughout,
$\=$ will stand for equality restricted to $\Delta$; an arrow under an
index will denote pull-back of that index to $\Delta$; $V^a$ will be a
generic vector field tangential to $\Delta$ and $\tilde{V}^a$ any of
its extensions to space-time. The electromagnetic potential and fields
will be denoted by bold-faced letters. All fields are assumed to be
smooth, and bundles, trivial. For details, generalizations and
subtleties, see \cite{abf,cs,afk,abl}.

\textsl{Definition:} A sub-manifold $\Delta$ of a space-time
$(M,g_{ab})$ is said to be an \textit{isolated horizon} if:\\
i) \textit{It is topologically $S^2\times R$, null, with zero shear and
expansion}. This condition implies, in particular, that the space-time
$\nabla$ induces a unique derivative operator $D$ on $\Delta$ via $D_a
V^b := \nabla\!_{\pback{a}} \tilde{V}^b$.\\
ii) $(\Lie_\l D_a - D_a\Lie_\l) V^b \= 0$ and $\Lie_{\l}
\emA\!_{\pback{a}} \= 0$ \textit{for} some \textit{null normal $\l$ to
$\Delta$}; and,\\
iii) \textit{Field equations hold at} $\Delta$.

All these conditions are \textit{local} to $\Delta$. The first two
imply that the intrinsic metric and connection on $\Delta$ are
`time-independent' and spell out the precise sense in which $\Delta$ is
`isolated'. Every Killing horizon which is topologically $S^2\times R$
is an isolated horizon. However, in general, space-times with isolated
horizons need not admit \textit{any} Killing field even in a
neighborhood of $\Delta$. The local existence of such space-times was
shown in \cite{l}. A global example is provided by Robinson-Trautman
space-times which admit an isolated horizon but have radiation in
\textit{every} neighborhood of it \cite{c}. Finally, on a general
$\Delta$, the null normal $\ell$ of ii) plays a role analogous to that
of the Killing field on a Killing horizon. Generically, $\l$ satisfying
ii) is unique up to a \textit{constant} rescaling $\l \rightarrow c
\l$. (In particular, this is true of the Kerr family.) We will denote
by $[\l ]$ the equivalence class of null normals satisfying ii). One
cannot hope to eliminate this constant rescaling freedom because,
without reference to infinity, it exists already on Killing horizons.

\textsl{Geometry of isolated horizons:} Although the boundary
conditions are rather weak, they have surprisingly rich consequences.
We now summarize the most important ones.\\
1. Intrinsic geometry: $\l$ is a symmetry of the degenerate, intrinsic
metric $q_{ab} := g_{\pback{ab}}$ of $\Delta$; $\Lie_\l q_{ab} \= 0$.
$\Delta$ is naturally equipped with a 2-form $\epsilon_{ab}$, the
pull-back to $\Delta$ of the volume 2-form on the 2-sphere of integral
curves of $\l$, satisfying $\epsilon_{ab}\l^b \= 0$ and $\Lie_\l
\epsilon_{ab} \= 0$. The area of any cross-section $S$ is given by
$\oint_S\, \epsilon$ and is the same for all cross-sections. We will
denote it by $a_\Delta$.\\
2. Connection coefficients: $\l$ is geodesic and free of divergence,
shear and twist. Hence there exists a 1-form $\omega$ on $\Delta$ such
that $\nabla\!\!_{\pback{a}}\l^b = \omega_a\l^b$. The surface gravity
$\kappa_{\ls}$ defined by $\l$ is given by $\kappa_{\ls} =
\omega_a\l^a$. The boundary conditions imply $\kappa_{\ls}$ is constant
on $\Delta$ \cite{afk}. Thus, the zeroth law holds. Similarly, the
electromagnetic potential $\Phi_{\ls} = -\emA_a\l^a$ is constant on
$\Delta$ \cite{afk}. Note, however, that other connection components or
the scalar curvature of the intrinsic metric $q_{ab}$ need not be
constant; the horizon may be distorted arbitrarily.\\
3. Weyl curvature: Let us pick an $\l$ in $[\l]$ and construct a null
tetrad $\l, n, m, \bar{m}$ on $\Delta$. Here $m,\bar{m}$ are chosen to
be tangential to $\Delta$ and thus $n$ is transverse. Then, the Weyl
components $\Psi_0 = C_{abcd} \l^a m^b \l^c m^d$ and $\Psi_1 = C_{abcd}
\l^a m^b \l^c n^d$ vanish, implying that there is no flux of
gravitational radiation across $\Delta$ and the Weyl tensor at $\Delta$
is of Petrov type II \cite{afk}. Hence $\Psi_2 := C_{abcd}\l^a m^b
\bar{m}^c n^d$ is gauge invariant. Its imaginary part is determined by
$\omega$ via: $d\omega = 2 \, {\rm Im}\, \Psi_2\, \epsilon$.  If there
are no matter fields \textit{on} $\Delta$, the horizon angular momentum
is determined entirely by ${\rm Im}\Psi_2$. While $\Psi_2$ is time
independent on the horizon, in general, $\Psi_3 =
C_{abcd}\l^am^b\bar{m}^cn^d$ and $\Psi_4 =
C_{abcd}n^a\bar{m}^bn^c\bar{m}^d$ are not \cite{abl}.\\
4. A natural foliation: Let us consider the non-extremal case when
$\kappa_{\ls}$ is non-zero. Then, $\Delta$ admits a natural foliation,
thereby providing a natural `horizon rest frame' \cite{abl}. The
2-sphere cross-sections of the horizon defined by this foliation are
completely analogous to the `good cuts' that null infinity admits in
absence of Bondi news. Therefore, we will refer to them as \textit{good
cuts} of the horizon. If there is no gravitational contribution to
angular momentum, i.e., if ${\rm Im}\,\Psi_2 \hateq 0$, then $d\omega$
vanishes. Hence, there exists a function $\psi$ on $\Delta$ with
$\omega \= d\psi$. Since $\Lie_\l \psi \= \omega\cdot \l \= \kappa$ is
constant on $\Delta$, the $\psi \= {\rm constant}$ surfaces foliate
$\Delta$. In the general case, the argument is more involved but the
foliation is again determined invariantly by the geometrical structure
of $\Delta$. This foliation turns out to be very useful (see below).\\
5.  Symmetries of $\Delta$: In view of our main Definition, it is
natural to define the symmetry group $\G$ of a given isolated horizon
to be the sub-group of ${\rm Diff}\, \Delta$ which preserves $[\l],\,
q_{ab},\, D, \emA\!_{\pback a}$.  Since $q_{ab},\, D ,\,
\bf{A}\!_{\pback a}$ can vary from one isolated horizon to another,
$\G$ is not canonical.  For simplicity, let us again restrict ourselves
to the non-extremal case $\kappa_{\ls}\not= 0$.  Then, isolated
horizons fall into three \textit{universality classes} \cite{abl}: I.
dim $\G$ = 4: in this case, $q_{ab}$ is spherically symmetric, good
cuts are invariant under the natural $SO(3)$ action and $\G$ is the
direct product of $SO(3)$ with translations along $\l$; II. dim $\G$ =
2: in this case, $q_{ab}$ is axi-symmetric, the general infinitesimal
symmetry $\xi^a$ has the form $\xi^a \= c \l^a + \Omega \varphi^a$,
where $c, \Omega$ are arbitrary constants on $\Delta$ and $\varphi$ is
a rotational vector field tangential to good cuts; and, III. dim $\G$
=1: in this case, the infinitesimal horizon symmetry has the form
$\xi^a = c \l^a$.  In case I, as one might expect, ${\rm Im}\, \Psi_2
\= 0$ and the horizon is non-rotating.  Case III corresponds to general
distortion.

\textsl{Extracting physics:} The isolated horizon framework can be used
to extract invariant physical information in the strong field region
near black holes, formed by gravitational collapse or merger of compact
objects.  At a sufficiently late time, the space-time would contain an
(approximate) isolated horizon $\Delta$.  In the most interesting case,
$\Delta$ would be of universality class II above. We will now focus on
this class and comment on other cases at the end of this letter. First,
we can ask for the angular momentum and mass of $\Delta$.  Recall that,
for asymptotically flat space-times without internal boundaries, one
obtains expressions of the ADM mass $M_\infty$ and angular momentum
$J_\infty$ using a Hamiltonian framework.  This strategy can be
extended to the present case (see below).  When constraints are
satisfied, the total Hamiltonian is now a sum of two surface terms, one
at infinity and the other at $\Delta$. The terms at infinity again
yield $M_\infty$ and $J_\infty$. General arguments lead one to
interpret the surface terms at $\Delta$ as the horizon mass $M_\Delta$
and angular momentum $J_\Delta$.  We have \cite{abl}:
\ba \label{J} J_\Delta &=& - \frac{1}{8\pi G}\oint_{S} \,
(\varphi \cdot \omega)  \, \epsilon +
2G (\varphi \cdot  \emA) \, {}^\star \emF \nonumber \\
&=& - \frac{1}{4\pi G}\oint_{S} \,
f ({\rm Im}\, \Psi_2 \, \epsilon + 2G\, {\rm Im}\, \phi_1\,
{}^\star \emF )\ea
where $S$ is any 2-sphere cross-section of $\Delta$, $f$ is related to
$\varphi$ by $D_a f = \epsilon_{ba }\varphi^b$ and ${\rm Im}\, \phi_1 =
-(i/2)\emF_{ab} m^a \bar{m}^b$ is a Newman-Penrose component of the
Maxwell field. In a vacuum, axi-symmetric space-time, $J_\Delta =
J_\infty$.  However, in general, the two differ by the angular momentum
in the gravitational radiation and the Maxwell field in the region
between $\Delta$ and infinity. Even in presence of such radiation, the
horizon mass is given by \cite{abl}
\be \label{M} M_\Delta = \frac{1}{2GR_\Delta}\left((R_\Delta^2+GQ^2)^2+
4G^2J_\Delta^2\right)^{\textstyle{1\over 2}} \ee
where $R_\Delta$ is the horizon radius, given by $a_\Delta = 4\pi
R_\Delta^2$, and $Q_\Delta = -\textstyle{1\over 4\pi}\oint_{S}
{}^\star\,\emF$ is the horizon charge. Somewhat surprisingly,
$M_\Delta$ has the same dependence on area, angular momentum and charge
as in the Kerr-Newman family (provided $J_\Delta$ is defined via
(\ref{J})).  However, this is a \textit{result} of the calculation, not
an assumption.  In a Kerr-Newman space-time, we have $M_\infty =
M_\Delta$ for all values of $Q$. (Thus, if $Q\not= 0$, $M_\Delta$ does
not agree with any of the known quasi-local expressions of mass.)
However, in general $M_\Delta$ is different from $M_\infty$.  Under
certain physically reasonable assumptions on the behavior of fields
near future time-like infinity $i^+$, one can show that the difference
is the energy radiated across ${\cal I}^+$ by gravitational and
electromagnetic waves.

If $\kappa_{\ls} \not=0$, irrespective of the universality class, one
can introduce (essentially) invariant coordinates and tetrads in a
neighborhood of $\Delta$. Fix an $\l$ in $[\l ]$.  Let $v,\theta,\phi$
be coordinates on $\Delta$ such that $\Lie_\l v \= 1$ and good cuts are
given by $v \= {\rm const}$. Let $n^a$ be the unique future-directed
null vector field which is orthogonal to the good cuts and normalized
so that $\l \cdot n \= -1$.  Consider past null geodesics emanating
from the good cuts, with $-n^a$ as their tangent at $\Delta$. Finally,
define $r$ via $\Lie_{n} r = -1$ and $r=r^o$ on $\Delta$, and Lie drag
$v,\theta,\phi$ along $n^a$. We now have a natural set of coordinates,
$(r,v,\theta,\phi)$, the only arbitrariness being in the
\textit{initial} choice of $(\theta, \phi)$ and adding constants to
$r,v$. Next, let us parallel transport $\l, m, \bar{m}$ along $n$ to
obtain a null tetrad in this neighborhood. The tetrad is unique up to
local $m$-$\bar{m}$ rotations \textit{at} $\Delta$. Now, assuming the
vacuum equations hold in this neighborhood, one can give a Bondi-type
expansion for the metric components in powers of $(r$-$r^o)$ to any
desired order. For example, retaining terms to second order, we have
\cite{abl}:
\[
\begin{array}{l@{\hspace{3pt}}l@{\hspace{3pt}}l}
  g_{ab}  &=& 2m^o_{(a}\bar{m}^o_{b)} + 2r_{,(a}v_{,b)}\ -\ (r-r^0)
      \big[ 4\mu^o m^o_{(a}\bar{m}^o_{b)}\\[1.5mm]
  &+& 2\lambda^o m^o_{(a}m^o_{b)} + 2\bar{\lambda}^o\bar{m}^o_{(a}\bar{m}^o_{b)} +
      2v_{,(a}(2\omega_{b)}- \kappa_{(\l)}v_{,b)})\big] \\[1.5mm]
  &+&  (1/2)(r-r^o)^2\big[ 4((\mu^o)^2 + \lambda^o\bar{\lambda}^o)
      m^o_{(a}\bar{m}^o_{b)} \\[1.5mm]
  &+& (4\mu^o\lambda^o - 2\Psi^o_4)m^o_{(a}m^o_{b)} + (4\mu^o\bar{\lambda}^o -
      2\bar{\Psi}^o_4)\bar{m}^o_{(a}\bar{m}^o_{b)}\nonumber\\[1.5mm]
  &+& 4(\bar{\pi}^o\lambda^o + \pi^o\mu^o - \Psi^o_3)v_{,(a} m^o_{b)}\\[1.5mm]
  &+& 4(\pi^o\bar{\lambda}^o + \bar{\pi}^o\mu^o
      - \bar{\Psi}^o_3) v_{,(a} \bar{m}^o_{b)}\\[1.5mm]
  &+& (2\pi^o \bar{\pi}^o -2\Psi^o_2 -2\bar{\Psi}^o_2)v_{,(a}v_{,b)}\big]
      + O(r-r^o)^3,
\end{array}
\]
where quantities with the a superscript $o$ are evaluated on $\Delta$,
and the Newman-Penrose spin coefficients are defined as: $\mu =
m^a\bar{m}^b\nabla_a n_b$, $\lambda = \bar{m}^a\bar{m}^b\nabla_an_b$
and $\pi = \l^a\bar{m}^b\nabla_an_b$. Using the boundary conditions and
field equations, at the horizon these spin coefficients as well as the
Weyl components can be expressed in terms of the dyad $m^o, \bar{m}^o$
defining the intrinsic horizon geometry, 1-form $\omega_a$ and the
value of $\Psi^o_4$ on any one good cut \cite{abl}.  The coefficient of
$(r - r^o)^n$ in the expansion is expressible in terms of these fields
\textit{and} the (n-2)th radial derivative of $\Psi_4$, evaluated on
$\Delta$.

The null surfaces $v= {\rm const.}$ are invariantly defined. Therefore
(modulo the small freedom mentioned above) the tetrad components of the
Weyl tensor on these surfaces are gauge invariant. This property will
be useful in physically interpreting the outcomes of numerical
simulations of mergers of compact objects. For example, it will enable
a gauge invariant comparison between the radiation fields $|\Psi_4|$
created in two simulations, say with somewhat different initial
conditions.  Finally, one can give a systematic procedure to extend any
infinitesimal symmetry $t^a \= c \l^a + \Omega \varphi^a$ on $\Delta$
to a `potential Killing field' $\tilde{t}^a$ in a neighborhood
\cite{abl}. If the space-time does admit a Killing field $\xi^a$ which
coincides with $t^a$ on $\Delta$, then $\xi^a$ must equal $\tilde{t}^a$
in the neighborhood. Again, since they are defined invariantly, the
vector fields $\tilde{t}^a$ can be useful to extract physical
information coded in the strong field geometry.

Finally, note that all this structure ---particularly the definitions
of $M_\Delta$ and $J_\Delta$--- is defined intrinsically, using local
geometry of the \textit{physical} space-time under consideration.  To
extract physical information, one does not have to embed this
space-time in a Kerr solution which, in the light of the no-hair
theorems, presumably approximates the physical, near horizon geometry
at late times.  In practice this is a significant advantage because the
embedding problem can be very difficult: typically, one knows little
about the form of the desired Kerr metric in the coordinate system in
which the numerical simulation is carried out.  More importantly, a
priori, one does not know which Kerr parameters to use in the
embedding, nor does one have a quantitative control on precisely how
the physical near-horizon geometry is to approach Kerr.

\textsl{Isolated Horizon Mechanics:} We already saw that the zeroth law
holds on all isolated horizons.  Let us consider the first law: $\delta
M = (\kappa/8\pi G) \delta a + \Omega \delta J + \Phi \delta Q$.  In
the stationary context the law is somewhat `hybrid' in that $M$ and $J$
are defined at infinity, $a$ at the horizon and $\kappa, \Omega$ and
$\Phi$ are evaluated at the horizon but refer to the normalization of
the Killing field carried out at infinity.  In the non-stationary
context now under consideration, there are two additional problems: due
to the presence of radiation, $M_\infty$ and $J_\infty$ have little to
do with the horizon mass and since we no longer have a global Killing
field, there is an ambiguity in the normalization of $\kappa$ and
$\Omega$.

As in \cite{byw}, our strategy is to arrive at the first law through a
Hamiltonian framework, but now adapted to the isolated horizon boundary
conditions.  For brevity, we will again focus on the physically most
interesting universality class II. Let us fix on the (abstract)
isolated horizon boundary $\Delta$ a rotational vector field
$\varphi^a$. Consider the space $\Gamma$ of asymptotically flat
solutions to the Einstein-Maxwell equations for which $\Delta$ is an
isolated horizon inner-boundary with symmetry $\varphi^a$. $\Gamma$
will be our covariant phase-space \cite{afk,abl}.  Denote by
$\tilde\varphi^a$ any extension of $\varphi^a$ which is an asymptotic
rotational Killing field at spatial infinity.  Then, one can show that
the vector field $\delta_{\tilde\varphi}$ on $\Gamma$ defined by the
Lie derivative of basic fields along $\tilde\varphi^a$ is a phase space
symmetry, i.e., Lie drags the symplectic structure. Its generator is
given by \cite{abl}
$$ H_{\tilde\varphi} = J_\infty - J_\Delta $$
where $J_\Delta$ is given by (\ref{J}). Hence, it is natural to
interpret (\ref{J}) as the horizon angular momentum.

To define the horizon energy, one needs to select a `time translation'.
On $\Delta$, it should coincide with a horizon symmetry $t^a \= c \l^a
+ \Omega \varphi^a$. While $c, \Omega$ are constants on $\Delta$, in
the phase space we must allow them to vary from one solution to
another.  (In the numerical relativity language, we must allow $t^a$
---or, the lapse and shift at $\Delta$--- to be \textit{live}.)  For,
unlike at infinity, the 4-geometries under consideration do not
approach a fixed 4-geometry at $\Delta$, whence it is not a priori
obvious how to pick the \textit{same} time-translation for all
geometries in the phase space. Let $\tilde{t}^a$ be any extension of
$t^a$ to the whole space-time which approaches a {\it fixed} time
translation at infinity.  We can ask if the corresponding
$\delta_{\tilde{t}}$ is a phase space symmetry.  The answer is rather
surprising: yes, \textit{if and only if there exists a function
$E^t_\Delta$ on the phase space, involving only the horizon fields,
such that the first law},
\be \delta E^t_\Delta = \frac{\kappa_{(t)}}{8\pi G}\, \delta a_\Delta +
\Omega_{t}\, \delta J_\Delta + \Phi_{(t)} \delta\, Q \, ,\ee
\textit{holds} \cite{afk,abl}. Thus, not only does the isolated horizon
framework enable one to extend the first law beyond the stationary
context, but it also brings out its deeper role: the first law is a
necessary and sufficient condition for a consistent Hamiltonian
evolution.

However, there are many choices of $t^a$ on the horizon for which this
condition can be met, each with a corresponding time-evolution, horizon
energy function and first law.  Can we make a canonical choice of
$t^a$?  In the Einstein-Maxwell theory, the answer is in the
affirmative.  The requirement that the (live) vector field
$\tilde{t}^a$ coincide, on each Kerr-Newman solution, with that
stationary Killing field which is unit at infinity \textit{uniquely}
fixes $t^a$ on the isolated horizon of every space-time in the phase
space.  With this canonical choice, say $t = t_o$, in Einstein-Maxwell
theory we can define the horizon mass to be
$$ M_\Delta = E^{t_o}_\Delta\, . $$
Then, $M_\Delta$ is given by (\ref{M}).

We will conclude with three remarks.\\
1. We focused our discussion on the physically most interesting
universality class II. Class I was treated in detail in \cite{abf,cs}
and is a special case of non-rotating, class III horizons discussed in
\cite{afk}. All these cases have been analyzed in detail.  However, the
current understanding of class III with rotation $({\rm Im}\, \Psi_2
\not= 0)$ is rather sketchy.\\
2.  The framework that led us to the zeroth and first laws can be
easily extended to other space-time dimensions.  The 2+1-dimensional
case has already been analyzed in detail \cite{adw} and has some
special interesting features in the context of a negative cosmological
constant.  In the non-rotating, class III case, dilaton and Yang-Mills
fields have also been incorporated \cite{abf,cs,afk}.  In the
Yang-Mills case, although the zeroth and first laws can be proved, the
analog of the mass formula (\ref{M}) is not known because one does not
have as much control on the space of all stationary solutions.
Nonetheless, the framework has been used to derive new relations
between masses of static black holes with hair and their solitonic
analogs in Einstein-Yang Mills theory\cite{cs,afk}. More importantly,
as is well-known, the standard no-hair theorems fail in this case and
the framework has been used to conjecture new no-hair theorems tailored
to isolated horizons rather than infinity \cite{cs}.\\
3.  In the non-rotating case, the framework has been used to carry out
a systematic and detailed entropy calculation using non-perturbative
quantum gravity \cite{abck}.  The analysis encompasses all black holes
without any restriction of near-extremality made in string theory
calculations.  Furthermore, it also naturally incorporates the
cosmological horizons to which thermodynamic considerations are known
to apply \cite{gh}.  Recently, sub-leading corrections to entropy have
also been calculated \cite{km}. However, the non-perturbative
quantization scheme faces a quantization ambiguity ---analogous to the
$\theta$-ambiguity in QCD--- which permeates all these calculations.
Its role is not fully understood. Carlip \cite{sc} and others have
suggested the use of horizon symmetries in entropy calculations and
this approach could shed light on the quantization ambiguity and relate
the analysis of \cite{abck} to conformal field theories. Conversely,
the isolated horizon framework may offer a more systematic avenue for
implementing Carlip's ideas. Finally, since rotation has now been
incorporated in the classical theory \cite{abl}, one can hope to extend
the entropy calculation to this case.

\textbf{Acknowledgements} We would like to thank A. Corichi, S.
Hayward, J. Pullin, D. Sudarsky and R. Wald for discussions. This work
was supported in part by the NSF grants PHYS95-14240, INT97-22514, the
Polish CSR grant 2 P03B 060 17, the Albert Einstein Institute and the
Eberly research funds of Penn State.

\end{document}